\documentclass[usedcolumn,usenatbib,usegraphicx]{mn2e}

\title[One-armed spiral structure of accretion discs around the neutron star in Be/X-ray binaries]
  {One-armed spiral structure of accretion discs induced by a phase-dependent mass transfer in Be/X-ray binaries}
\author[Kimitake~Hayasaki and Atsuo~T.~Okazaki]
  {Kimitake~Hayasaki$^1$$^,$$^2$\thanks{E-mail: hayasaki@topology.coe.hokudai.ac.jp}
   and Atsuo~T.~Okazaki$^3$ \\
  $^1$Department of Applied Physics, Graduate School of Engineering, 
	Hokkaido University, Kitaku N13W8, Sapporo 060-8628, Japan.\\
  $^2$Centre for Astrophysics and Supercomputing, Swinburne University of Technology, 
	Hawthorn Victoria 3122 Australia.\\ 
  $^3$Faculty of Engineering, Hokkai-Gakuen University, Toyohira-ku, 
      Sapporo 062-8605, Japan.}
\date{Received ...,; accepted ...}

\pagerange{\pageref{firstpage}--\pageref{lastpage}} \pubyear{2004}

\def\LaTeX{L\kern-.36em\raise.3ex\hbox{a}\kern-.15em
    T\kern-.1667em\lower.7ex\hbox{E}\kern-.125emX}

\begin{document}

\label{firstpage}

\maketitle

\begin{abstract}

We study non-axisymmetric structure of accretion discs
in Be/X-ray binaries, performing three dimensional Smoothed Particle
Hydrodynamics simulations for a coplanar system with a short period
and a moderate eccentricity.
We find that ram pressure due to the phase-dependent mass transfer
from the Be-star disc excites a one-armed, trailing spiral structure 
in the accretion disc around the neutron star. 
The spiral wave is transient; it is excited around the periastron
passage, when the material is transferred from the Be disc, and
is gradually damped afterwards. 
The disc changes its morphology from circular to eccentric 
with the development of the spiral wave, 
and then from eccentric to circular with the decay of the wave 
during one orbital period.
It turns out that the inward propagation of the spiral wave
significantly enhances the mass-accretion rate onto the neutron star.
Thus, the detection of an X-ray luminosity peak 
corresponding to the peak in enhanced mass-accretion rate
provides circumstantial evidence that 
an accretion disc is present in Be/X-ray binaries.

\end{abstract}


\section{Introduction}
\label{sec:intro}

About two-thirds of high-mass X-ray binaries
have been identified as Be/X-ray binaries. 
These systems generally consist of a neutron star and
a Be star with a cool ($\sim 10^{4}K$) equatorial disc, which is
geometrically thin and nearly Keplerian. 
Be/X-xay binaries are distributed over a wide range of orbital
periods $(10\,{\rm d} \la P_{\rm{orb}} \la 300\,\rmn{d})$ and
eccentricities ($e \la 0.9$).

Most Be/X-ray binaries show only transient activity in 
X-ray emission and are termed Be/X-ray transients. Be/X-ray transients
show periodic (Type I) outbursts, which are separated by the orbital
period and have luminosity 
$L_{\rmn{X}}=10^{36-37}\rmn{erg\,s}^{-1}$, and giant (Type II)
outbursts of $L_{\rmn{X}} \ga 10^{37} \rmn{erg\,s}^{-1}$ with no
orbital modulation. 
These outbursts have features that strongly suggest the presence of
an accretion disc around the neutron star.

Recently, Hayasaki \& Okazaki (2004, hereafter paper~I) studied the
accretion flow around the neutron star in a Be/X-ray binary with a
short period ($P_{\rm orb}=24.3\,{\rm d}$) and a moderate eccentricity
($e=0.34$), using a three dimensional Smoothed Particle Hydrodynamics
(SPH) code.
They found that a time-dependent accretion disc is formed around the
neutron star. They also discussed the evolution of the
azimuthally-averaged structure of the disc, although the disc is 
significantly eccentric in spite of the nealry Keplerian rotation.
The non-axisymmetric structure seen in the disc is considered 
to result from a phase-dependent mass transfer from the Be disc.

\citet{tosa} showed numerically that ram pressure 
due to an intergalactic gas excites the one-armed spiral structure in
the non-selfgravitating, galactic $V$-constant disc. 
Then \citet{ka2} showed analytically that
the excitation of the one-armed mode is caused by the
same resonant instability mechanism as
the tidally induced eccentric instability which explains
the superhump phenomena in dwarf novae systems (\citealt{hi}, \citealt{lu}).
In the context of accretion disc theory,  
\citet{lu1} investigated the effect of time varying mass-transfer 
on the eccentricity of an accretion discs in a circular binary. 
However, it has never been explicitly studied 
how phase-dependent mass transfer affects 
the eccentricity of an accretion disc in an eccentric binary.

In this {\it letter}, we study the non-axisymmetric structure of 
the accretion disc around the neutron star in Be/X-ray binaries.
In Section~\ref{sec:ram}, the development of the 
one-armed spiral structure is demonstrated with 3D SPH simulations.
The strength of several modes excited in the disc, 
the orbital modulation of the disc radius and 
the phase dependence of the mass-accretion rate are also analyzed 
in this section. Our conclusions are presented in Section~\ref{sec:conc}.

\section{Ram-pressure-deformed accretion discs}
\label{sec:ram}

In this section, we show that 
ram pressure due to the material transferred from the
Be disc around periastron temporarily excites the one-armed spiral wave 
in the accretion disc around the neutron star in Be/X-ray binaries.

Our simulations were performed using the same 3D SPH code as in
paper~I, which was based on a version originally developed by Benz
(Benz 1990; Benz et al.\ 1990) and later by Bate, Bonnell \& Price
(1995). In order to investigate the effect of ram pressure on the
accretion disc, we compare results from model~1 in paper~I (hereafter,
model~A) with those from a simulation (hereafter, model~B) in which
the mass transfer from the Be disc is artificially stopped for one
orbital period. Except for this difference, the two simulations have the
same model parameters: The orbital period $P_\rmn{orb}$ is
24.3\,d, the eccentricity $e$ is 0.34, and the Be disc is coplanar
with the orbital plane. The inner radius of the simulation region
$r_{\rm{in}}$ is $3.0\times10^{-3}a$, where $a$ is the semi-major axis
of the binary. A polytropic equation of state with the exponent 
$\Gamma=1.2$ is adopted. 
The Shakura-Sunyaev viscosity parameter
$\alpha_{\rm{SS}}=0.1$ throughout the disc. 
There is also a bulk viscosity
$\nu_{b}=5\alpha_{\rm{SS}}c_{\rm{s}}H/3$, 
where $c_{s}$ is the polytropic sound speed 
and $H$ is the scale-hight of the disc. 

\begin{figure}
\resizebox{\hsize}{!}
{
\includegraphics*{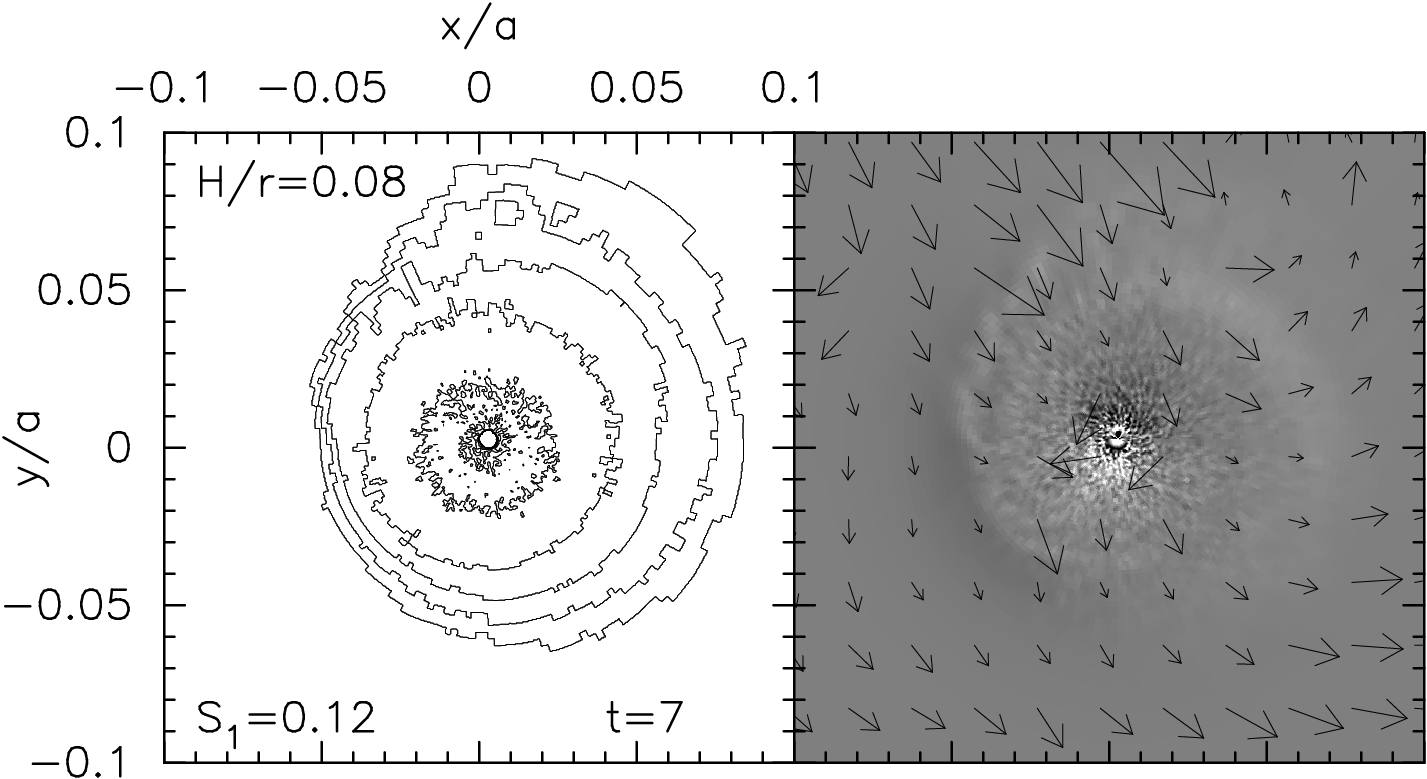}}\\
\resizebox{\hsize}{!}
{
\includegraphics*{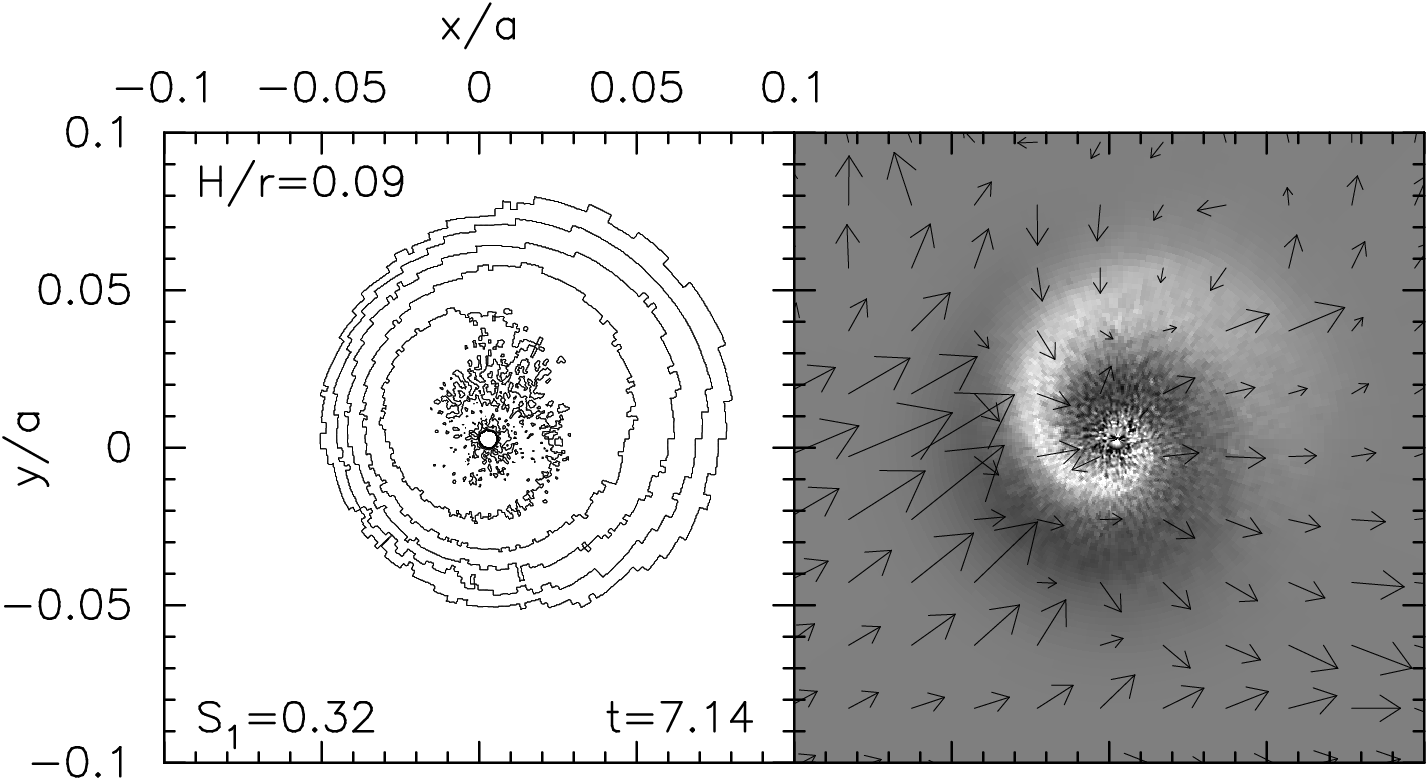}}\\
\resizebox{\hsize}{!}
{
\includegraphics*{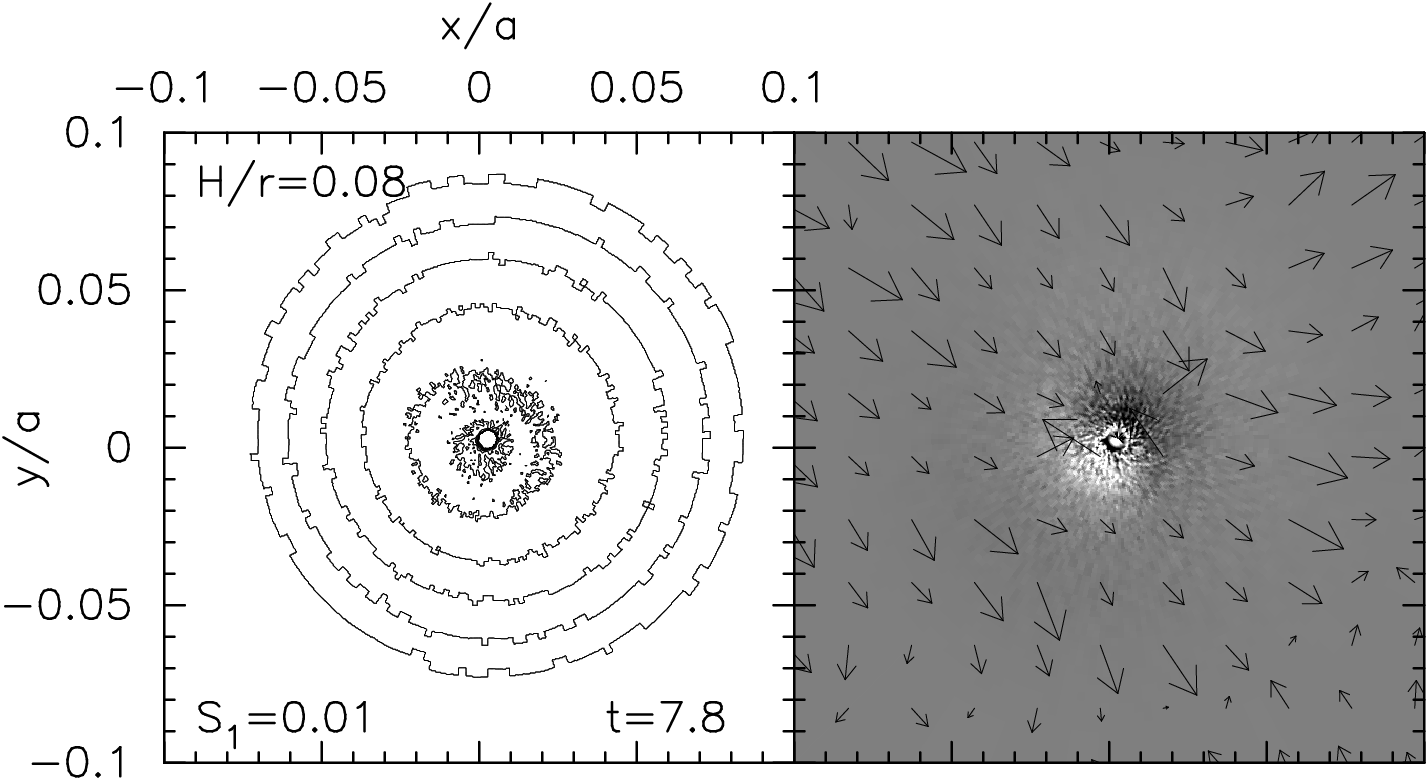}}
 \caption{
   Snapshots of the accretion disc for model~A.
   The left panels show the surface density in a range of 
   three orders of magnitude in the logarithmic scale,
   while the right panels show the non-axisymmetric components of 
   the surface density (gray-scale plot)
   and the velocity field (arrows) in the linear scale.
   In the right panels, the region in gray (white) denotes the region 
   with positive (negative) density enhancement.
   The periastron is in the $x$-direction and
   the disc rotates counterclockwise.
   Annotated in each left panel are the time 
   in units of $P_\rmn{orb}$, the relative disc thickness $H/r$ at $r=0.05a$ 
   and the mode strength $S_{1}$.} 
 \label{fig:contours}
\end{figure}

Fig.~\ref{fig:contours} gives a sequence of snapshots of the accretion
disc around the neutron star for $7\le{t}\le8$ in model~A, 
where the unit of time is the orbital period $P_\rmn{orb}$.
The left panels show the contour maps of the surface density, 
whereas the non-axisymmetric components of the surface density 
and the velocity field are shown in the right panels.
Annotated in each left panel are the time in units of $P_\rmn{orb}$, 
the relative disk thickness $H/r$ at $r=0.05\rm{a}$ 
and the mode strength $S_{1}$, a measure of the amplitude of 
the one-armed spiral wave,
details of which will be described later.
It is noted from the figure
that the one-armed, trailing spiral is excited at periastron 
and is gradually damped towards the next periastron.

\begin{figure}
\resizebox{\hsize}{!}
{
\includegraphics*{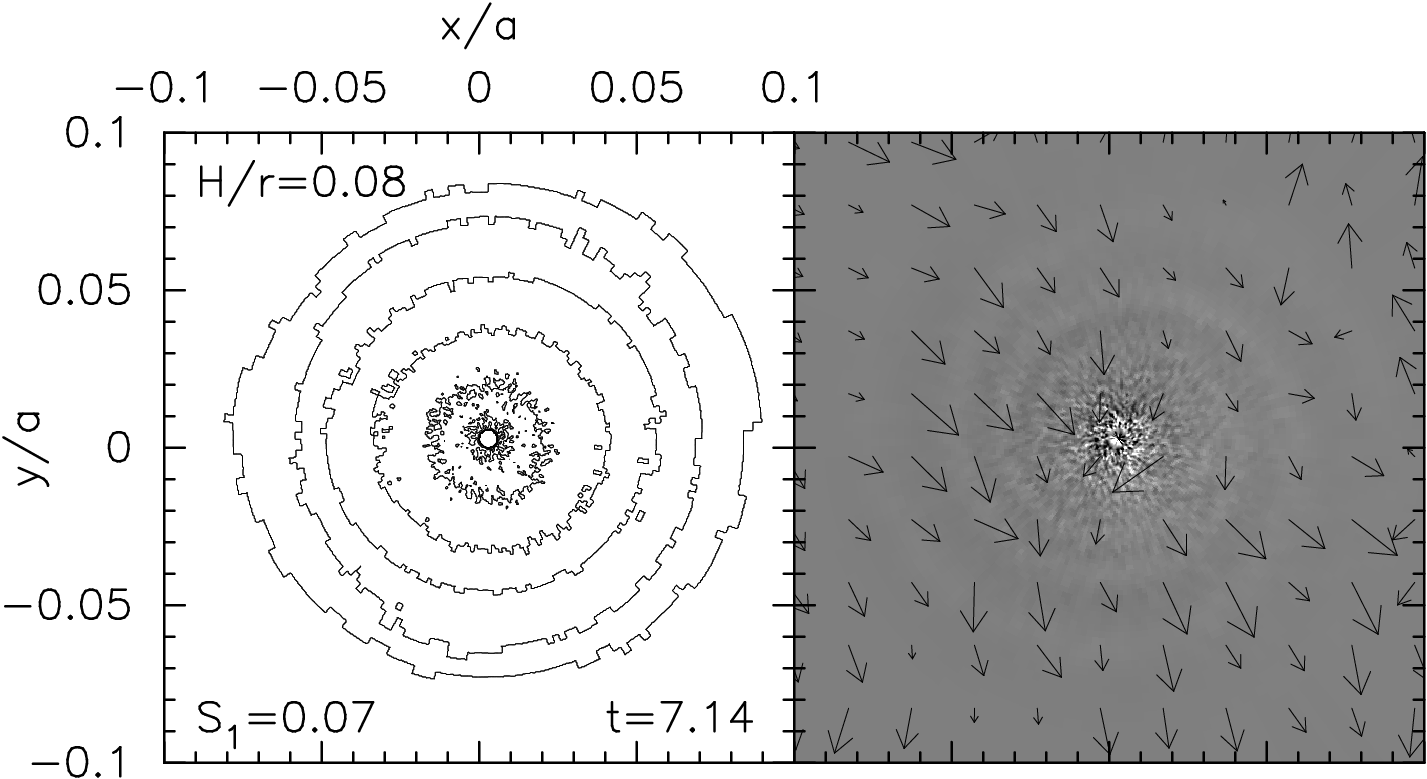}}\\
 \caption{Same as Fig.~\ref{fig:contours}, but for model~B.}
 \label{fig:contours2}
\end{figure}

For comparison, we present the results for model~B, in which
the mass transfer is artificially turned off for $7 \le t \le 8$.
Fig.~\ref{fig:contours2} shows 
the surface density (the left panel) 
and the non-axisymmetric components of the surface density 
and the velocity field (the
right panel) at the time corresponding to the middle panel of
Fig.~\ref{fig:contours}. The format of the figure is the same as that of
Fig.~\ref{fig:contours}.
It should be noted that 
the disc deformation due to the one-armed mode is not seen in model~B.
The disc is more circular and has a larger radius in model~B than in
model~A.
This strongly suggests that 
the excitation of the one-armed spiral structure in the accretion disc
is induced by ram pressure from the material transferred from the
Be disc at periastron.

\subsection{Mode strength}
\label{sec:stmode}

In this subsection, we analyse the orbital-phase dependence of 
the disc eccentricity 
by decomposing the surface density distribution 
into Fourier components which vary as $\exp{(im\phi)}$, where $m$ is 
the azimuthal harmonic number.

Following \citet{lu}, we define the mode strength by
\begin{equation}
S_{f,m}(t)\equiv\frac{2}{M_{d}(1+\delta_{m,0})}\int_{r}dr\int_{0}^{2\pi}rd\phi\Sigma(r,\phi,t)f(m\phi),
\label{eq:mode1}
\end{equation}
where $f$ is either $\sin$ or $\cos$ and
$M_{d}$ is the total disc mass given by $\int_{r}dr\int_{0}^{2\pi}rd\phi\Sigma(r,\phi,t)$.
Then, the total strength of the mode $m$ is defined by
\begin{equation}
S_{m}(t) = (S_{\cos, m}^{2} + S_{\sin, m}^{2})^{1/2}.
\label{eq:mode2}
\end{equation}

Note that our definition of the mode strength slightly differs 
from that of \citet{lu}. He decomposed the surface density distribution
into double Fourier components with the azimuthal and time harmonic numbers,
whereas we did it with only the azimuthal harmonic number. This is because
our purpose is to analyze the time dependence of the non-axisymmetric structure 
of the accretion disc in detail.
Note also that the definition of the mode strength is not unique,
so Eq.~(\ref{eq:mode2}) gives only a rough measure of the amplitude
of each mode. If the spiral wave is tightly winding,
Eq.~(\ref{eq:mode2}) will underestimate the amplitude significantly.

\begin{figure}
\resizebox{\hsize}{!}
{
\includegraphics*[angle=0]{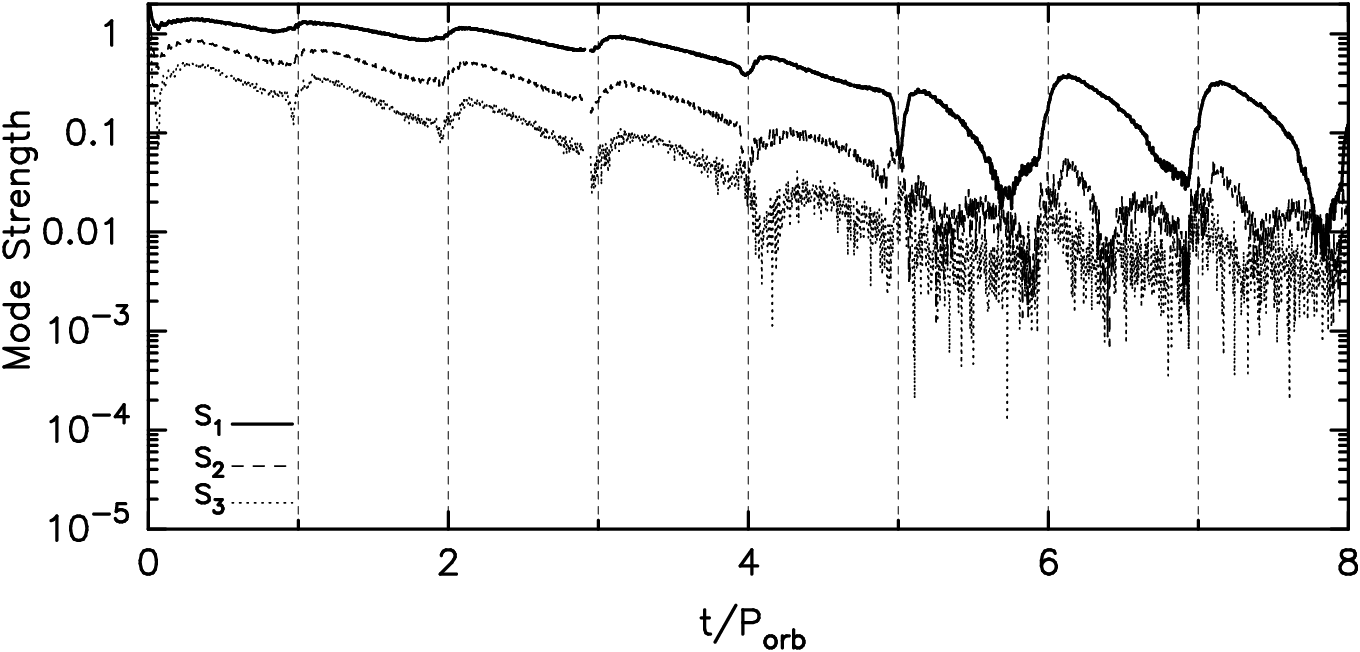}}
 \caption{
Evolution of several non-axisymmetric modes for $0\le{t}\le8$. 
The soild, dashed and dotted lines denote 
the strengths of the $m=1$, $m=2$ and $m=3$ modes, respectively.}
 \label{fig:mode}
\end{figure}

Fig.~\ref{fig:mode} shows the evolution of several non-axisymmetric modes
for model~A, in which the phase-dependent mass transfer from the Be disc 
is taken into account.
The solid, dashed and dotted lines denote the strengths 
of $m=1$, $m=2$ and $m=3$ modes, respectively.
From Fig.~\ref{fig:mode}, 
we note that the $m=1$ component dominates the other components 
throughout the run.
In the initial developing stage of the disc ($0\le{t}\le5$),
elliptical orbits of the gas particles transferred from the Be disc
significantly contribute to the strength of the $m=1$ component, $S_1$.
The contribution becomes smaller as the disc develops. Thus,
the decrease in $S_1$ for $t \la 5$ corresponds to the decrease in
the disc eccentricity during this period.
The decrease in eccentricity is likely to result from a non-zero bulk 
viscosity which helps to prevent  the spontaneous growth of the eccentricity
through viscous overstability \citep{ogi}.

In contrast, after the disc is well developed ($t \ga 5$),
the one-armed spiral wave gives the major contribution to $S_1$.
The mode strength significantly modulates with orbital phase,
corresponding to the rapid excitation and the subsequent gradual damping
of the wave.

\begin{figure}
\resizebox{\hsize}{!}
{
\includegraphics*[height=3cm,angle=0]{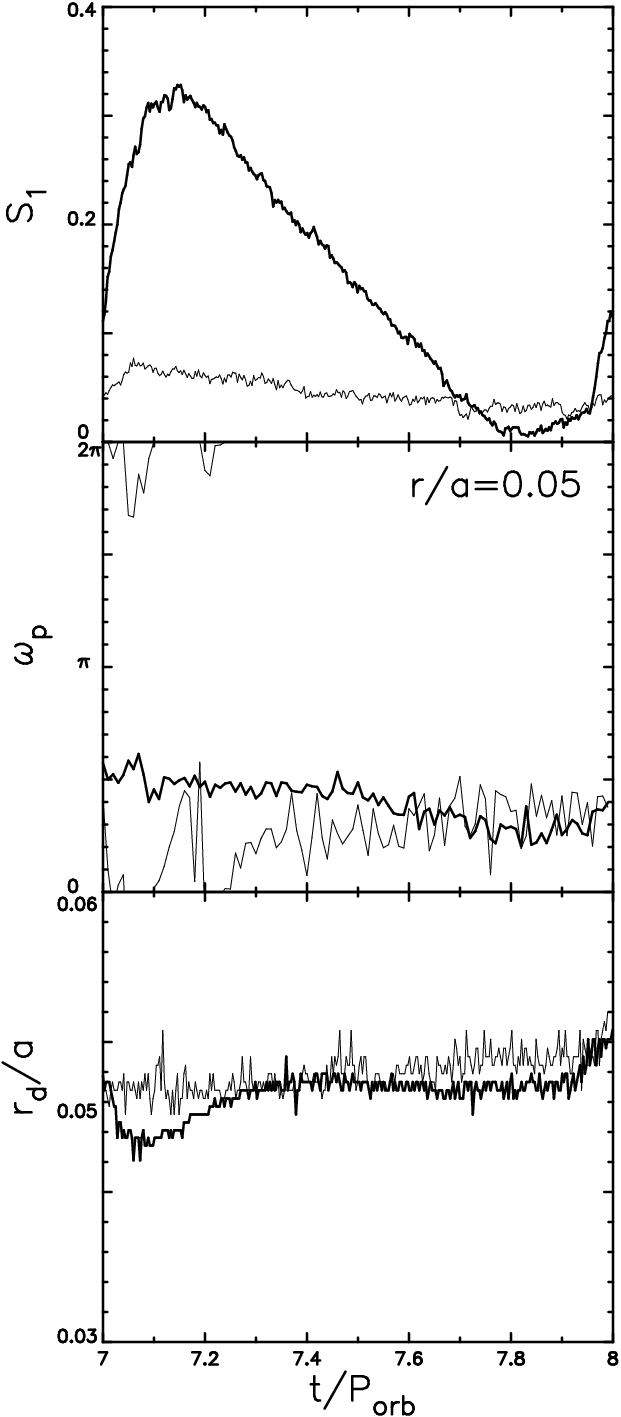}
}

 \caption{Time dependence of the mode strength $S_1$, 
          the azimuth of the maximum density perturbation, $\omega_{\rm{p}}$,
          at $r=0.05a$,
          and the disc radius $r_\rmn{d}$ for $7\le{t}\le8$.
          In each panel, 
          the thick line is for model~A, while the thin line for model~B.}
 \label{fig:one-armed}
\end{figure}

In order to see the time dependence of 
non-axisymmetric structure in the developed accretion disc more closely,
we show
the time dependence of the mode strength $S_1$ for $7\le{t}\le8$
 in the top panel of Fig.~\ref{fig:one-armed}.
The thick and thin lines denote $S_1$ in model~A and in model~B,
respectively. Recall that in model~B the mass transfer from the Be disc 
is artificially turned off for this period.

In model~A, the mode strength $S_1$ is dominated by the one-armed spiral mode.
It rapidly increases to the maximum value of $\sim 0.3$ at $t=7.14$
(see the middle panel of Fig.~\ref{fig:contours})
and then gradually decreases to a minimum value of $\sim 0.01$ at $t=7.8$
(the bottom panel of Fig.~\ref{fig:contours}).
It is important to note that the mode strength $S_1$ in model~A 
is much stronger and has much larger orbital modulation than in model~B.
This clearly shows that the one-armed mode is not excited by 
the tidal interaction with the Be star but by ram pressure of
the material transferred from the Be disc at periastron passage.

The middle panel of Fig.~\ref{fig:one-armed} shows 
the azimuth $\omega_{\rm{p}}$ between the position vector of 
the maximum value of the non-axisymmetric component of 
the surface density and the eccentric vector of the binary orbit 
(hereafter, the azimuth of the maximum density perturbation).
As in the other panels, the thick and thin lines are for  
model~A and model~B, respectively.
It should be noted that $\omega_\rmn{p}$ in both models shifts little
over one orbital period. This is a typical characteristic of the one-armed
modes in nearly Keplerian discs \citep{ka1}.

\subsection{Orbital modulation in the disc radius}
\label{sec:discsize}

In paper~I, we found that
the disc size modulates with the orbital phase; the disc shrinks 
at periastron and gradually restores its radius afterwards.
Our interpretation of this orbital modulation was that the decrease in  
the disc radius at periastron is due to a negative torque by the Be star, 
while the later expansion is due to the viscous diffusion.
For the reason described below, this interpretation must be corrected.

The bottom panel of Fig.~\ref{fig:one-armed} shows
the time dependence of the disc radius for $7\le{t}\le8$, which is defined in
the same way as in paper~I (equation 5). 
As in the other panels, the thick and thin lines are for model~A and model~B, 
respectively.
From the figure, we note that the disc radius in model~B 
monotonically increases with time and shows no orbital modulation.
This is in sharp contrast to the orbital modulation
in the disc radius seen in model~A. 
Since the only difference between model~A and model~B is that 
the mass transfer from the Be disc is included in model~A, 
while it is not in model~B, it is most likely that 
the reduction in the disc radius after periastron passage is related to
the mass transfer from the Be disc. 
Apparently, the negative torque by the Be star
has little effect on the accretion disc radius in these models.

How can the mass transfer from the Be star reduce 
the accretion disc size?
As shown in Fig.~3 of paper~I,
the circularization radius of the material transferred from the Be disc 
is smallest around the periastron.
It is much smaller than the accretion disc radius shown 
in the bottom panel of Fig.~\ref{fig:one-armed}.
In other words, the specific angular momentum of the material transferred
from the Be disc is much smaller than that of the material at the outer
radius of the accretion disc.
Accretion of such material onto the disc outer radius will temporarily
reduce the disc radius. This mechanism, together with the viscous diffusion,
naturally explains the orbital modulation in the disc radius found in paper~I.


\subsection{Phase dependence of the accretion rate}
\label{sec:accrate}

After the accretion disc is developed ($t \ga 5$), 
the mass-accretion rate has double peaks per orbit,
a relatively-narrow, low peak at periastron and a broad, high peak afterwards 
[see Fig.~15(a) of paper~I].
While the first low peak at periastron could be artificial, being related to 
the presence of the inner simulation boundary,
the origin of the second high peak was not clear.
Below we show that the one-armed spiral wave is responsible for the second peak
in the mass-accretion rate.

\begin{figure}
\resizebox{\hsize}{!}
{
\includegraphics*{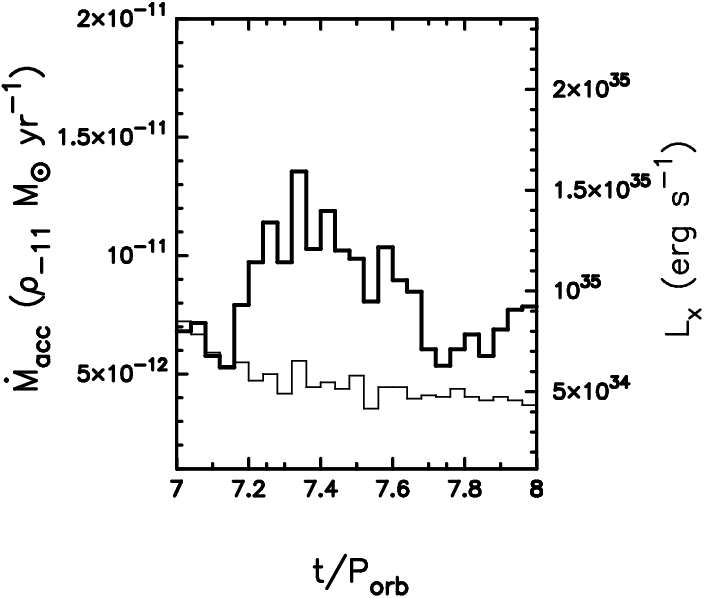}}\\
 \caption{Time dependence of the mass-accretion rate for $7\le{t}\le8$.
The thick and thin lines are for model~A and model~B, respectively. 
The right axis shows the X-ray luminocity corresponding 
to the mass-accretion rate.}
 \label{fig:accrate}
\end{figure}

Fig.~\ref{fig:accrate} shows 
the time dependence of the mass-accretion rate for $7\le{t}\le8$.
The thick line denotes the mass-accretion rate in
model~A, in which the mass transfer from the Be disc is taken into account.
For comparison, the mass-accretion rate in model~B, 
in which the mass transfer from the Be disc is artificially 
turned off at $t=7$, is also shown by the thin line. 
The difference between the accretion rate profiles for these two models is striking.
The accretion rate in model~B monotonically decreases over one orbital period,
whereas that of model~A shows a broad peak centred at $t\sim7.32-7.35$,
which corresponds to the second peak found in paper~I.

Although it is obvious that the above peak is caused by the mass transfer 
from the Be disc,
the mass transfer rate has a narrow peak at periastron. 
What mechanisms connecting the mass transfer from the Be disc 
and the mass accretion onto the neutron star
lead to a phase delay of $\sim 0.3$? 
There are two possibilites. 
Either the disc is the viscously adjusting to the addition 
of low-angular momentum material. However, even if the angular momentum 
of the material transferred from the Be disc is the lowest, 
its viscous time scale at the circularization radius 
is over 7 binary periods [see Fig.~3 of paper~I]. 
Thus, the viscous accretion of the lowest-angular momentum material 
could play no key role of the enhancement of the accretion rate at 
an orbital phase of about 0.35. 
Or, the second possibility is that the one-armed spiral wave is excited
by ram pressure of the transferred material from 
the Be disc and then the spiral wave travels inwardly.
The excitation of the spiral wave takes about $\sim (0.1-0.2) P_\rmn{orb}$ 
(see the top panel of Fig.~\ref{fig:one-armed}) and 
the difference of the peak position on the orbital phase 
between the mode strength and the accretion rate 
results from the wave propagation 
from the disc outer radius to the inner simulation boundary.

The time-scale for wave propagation is roughly estimated by
using the dispersion relation of the one-armed oscillation 
in the nearly Keplerian discs \citep{ka1}. Its frequency 
$\omega$ is written by $\omega{\sim}-\Omega_{\rm{orb}}(k_{\rm{r}}H)^{2}/2$, 
where $k_{\rm{r}}$ is the radial wavenumber and 
$\Omega_{\rm{orb}}$ is the orbital angular velocity. 
Finally, the time-scale for a global, one-armed pertubation to travel across the disc
is $2\pi/\omega\sim0.13P_{\rm{orb}}$ at $r\sim0.05a$ at $t\sim7.1$.
Therefore, combining the time-scales for wave excitation and propagation, 
we expect that the peak of the 
mass-accretion rate lags behind that of the mass-transfer rate 
by $(0.2-0.3) P_\rmn{orb}$,
which is in good agreement with the phase of the second peak 
of mass-accretion rate shown in Fig.~\ref{fig:accrate}.

\section{conclusions}
\label{sec:conc}

In this {\it letter}, 
we have investigated non-axisymmetric structure 
in the accretion disc around the neutron star in Be/X-ray binaries, 
analysing the results from a simulation performed by \citet{haya}.
We have adopted the phase-dependent, mass-transfer rate 
from a high-resolution simulation by 
\citet{oka2} for a coplanar system with a short period
($P_{\rmn{orb}}=24.3\,\rmn{d}$) and a moderate eccentricity $(e=0.34)$,
which targeted 4U\,0115+63, one of the best studied Be/X-ray
binaries. 

We have found that a one-armed, trailing spiral wave is excited in 
the accretion disc by ram pressure
from the strongly phase-dependent mass-transfer from the Be disc.
The spiral wave is transient; it is excited at periastron
passage, when mass transfer from the Be disc takes place, and
is gradually damped afterwards.
The one-armed perturbation pattern in the accretion disc precesses little 
over one orbital period, which is consistent with the theory of global, 
one-armed oscillations in nearly Keplerian discs \citep{ka1}.

The eccentric ($m=1$) mode dominates the other non-axisymmetric modes 
throughout the run ($0\le{t}\le8 P_\rmn{orb}$). 
In the initial stage of disc development ($0\le{t}\le5 P_\rmn{orb}$),
elliptical orbits of the gas particles transferred from the Be disc
make the accretion disc significantly eccentric.
After the disc is well developed ($t \ga 5$), however,
the one-armed spiral wave gives the major contribution to 
the eccentric deformation of the accretion disc.
Then, the strength of the deformation significantly modulates 
with the orbital phase,
corresponding to the rapid excitation and the subsequent 
gradual damping of the one-armed wave.

The system we studied differs from that of \citet{ka2}
in the sense that our accretion disc is nearly Keplerian and
ram pressure of the material transferred from the Be disc 
exerts on the accretion disc only briefly around periastron.
Nevertheless, we can estimate the growth rate of 
the one-armed spiral wave based on their formalism.
According to \citet{ka2},
the growth rate $\lambda$ is given by $\lambda\simeq
\pi{v_{r}}/(r_{\rm{d}}-r_{\rm{in}})$ where $v_{r}$ is 
the radial velocity of the disc. 
In our case, adopting $v_r \sim 0.07 a \Omega_\rmn{orb}$ 
at $r \sim r_\rmn{d} \sim 0.05a$, we have 
$\lambda\sim\Omega_{\rm{K}} \gg \Omega_\rmn{orb}$ at 
$r=0.05a$ for $7.0\le{t}\le7.2$.
This explains the rapid growth of the spiral wave
for $7.0\le{t}\le7.2$ in our simulation.

After the accretion disc has developed ($t \ga 5 P_\rmn{orb}$), 
the mass-accretion rate has double peaks per orbit,
a relatively-narrow, low peak at periastron and 
a broad, high peak afterwards. 
We have found that the one-armed spiral wave excited by ram pressure 
of the material transferred from the Be disc enhances the inward mass flux 
in the accretion disc, and is therefore responsible for this 
high peak in the accretion rate.
The X-ray luminosity peak corresponding to the high peak 
in accretion rate has never been identified in Be/X-ray binaries. 
However, if this peak is detected, 
it could be understood as strong circumstantial evidence 
for the presence of an accretion disc
in Be/X-ray binaries.

In this {\it letter}, we have reported on the effects of ram pressure 
on the accretion disc, including the excitation of the one-armed 
spiral wave, which enhances the accretion rate.
These results are based on the simulations, 
of which the run time was much shorter than the viscous time-scale.
In longer simulation runs, new features/interactions are expected to appear.
While the deformation of the accretion disc by ram pressure
is expected to become small as the disc mass increases,
the excitation of the eccentric mode directly driven by the one-armed 
bar potential, which is not seen in our simulations, will become important.
The tidal/resonant torque by the Be star may begin to work on 
the outer part of the accretion disc, as the disc size increases.
In a subsequent paper, we will investigate the long-term evolution of 
the accretion disc in Be/X-ray binaries in detail.

\section*{Acknowledgements}
We acknowlege the anonymous refree for constructive comments.
KH thank James R. Murray for useful comments.
The simulations reported here were performed using the facility at the
Hokkaido University Information Initiative Center, Japan.
This work has been supported by Grant-in-Aid for the 21st Century 
COE Scientific Research Programme on "Topological Science and Technology" 
from the Ministry of 
Education, Culture, Sport, Science and Technology of Japan (MECSST) and
in part by Grant-in-Aid for Scientific Reserch
(15204010 and 16540218) of Japan Society for the Promotion of Science.

\appendix

\end{document}